# Current status of the Spectrograph System for the SuMIRe/PFS

S. Vives *[a], D. Le Mignant[a], J.E. Gunn[b], S. Smee[c], L. Souza de Oliveira[d], N. Tamura[e], H. Sugai[e], R. Barkhouser[c], A. Bozier[a], M.A. Carr[b], A. Cesar de Oliveira[d], D. Ferrand[a], M. Golebiowski[c], M. Hart[c], S. Hope[c], M. Jaquet[a], F. Madec[a], S. Pascal[a], T. Pegot-Ogier[a], M. Vittal de Arruda[d]

[a]Aix Marseille Université, CNRS, LAM (Laboratoire d'Astrophysique de Marseille) UMR 7326, 13388, Marseille, France
[b]Department of Astrophysical Sciences, Princeton University, Princeton, NJ 08544, USA
[c]Department of Physics and Astronomy, Johns Hopkins University, Baltimore, MD, USA
[d]Laboratorio Nacional de Astrofisica/MCTI, 37504-364 Itajubá, Brasil
[e]Institute for the Physics and Mathematics of the Universe (IPMU), University of Tokyo (Japan)

## ABSTRACT

The Prime Focus Spectrograph (PFS) is a new facility instrument for Subaru Telescope which will be installed in around 2017. It is a multi-object spectrograph fed by about 2400 fibers placed at the prime focus covering a hexagonal field-of-view with 1.35 deg diagonals and capable of simultaneously obtaining data of spectra with wavelengths ranging from 0.38 um to 1.26 um.

The spectrograph system is composed of four identical modules each receiving the light from 600 fibers. Each module incorporates three channels covering the wavelength ranges 0.38–0.65 mu ("Blue"), 0.63–0.97 mu ("Red"), and 0.94–1.26 mu ("NIR") respectively; with resolving power which progresses fairly smoothly from about 2000 in the blue to about 4000 in the infrared. An additional spectral mode allows reaching a spectral resolution of 5000 at 0.8mu (red). The proposed optical design is based on a Schmidt collimator facing three Schmidt cameras (one per spectral channel). This architecture is very robust, well known and documented. It allows for high image quality with only few simple elements (high throughput) at the expense of the central obscuration, which leads to larger optics.

Each module has to be modular in its design to allow for integration and tests and for its safe transport up to the telescope: this is the main driver for the mechanical design. In particular, each module will be firstly fully integrated and validated at LAM (France) before it is shipped to Hawaii. All sub-assemblies will be indexed on the bench to allow for their accurate repositioning.

This paper will give an overview of the spectrograph system which has successfully passed the Critical Design Review (CDR) in 2014 March and which is now in the construction phase.

## 1. INTRODUCTION

The Prime Focus Spectrograph (PFS) of the Subaru Measurement of Images and Redshifts (SuMIRe) project [1] has been endorsed by the Japanese community as one of the main future instruments of the Subaru 8.2-meter telescope at Mauna Kea, Hawaii. PFS is a multiplexed fiber-fed optical and near-infrared spectrograph (Nfiber=2394), offering unique opportunities in survey astronomy and targeting cosmology with galaxy surveys, Galactic archaeology, and studies of galaxy/AGN evolution [2].

PFS is composed on 4 main elements: the Prime Focus Instruments (PFI) which covers a field-of-view of 1.3deg with 2400 fiber positioners, the fiber system which relays the light captured by each PFI positioner to the Spectrograph System (SpS), described in this paper. The last element is the metrology camera (located at the Cassegrain focus of the telescope) which will verify the correct location of each PFI fiber positioner.

*sebastien.vives@lam.fr; phone +33 (0)491 056 931

This article gives the current status of the opto-mechanical design of the Spectrograph System (SpS).

The Spectrograph System (SpS) is composed of four identical modules fed by 600 fibers each. Each module incorporates three channels covering the wavelength ranges 0.38-0.65μm ("Blue"), 0.63-0.97μm ("Red"), and 0.94-1.26μm ("NIR") respectively; with resolving power which progresses fairly smoothly from about 2000 in the blue to about 5000 in the infrared. In the Red channel, the spectral resolution can be increase up to 5000 by exchanging the disperser. Table 1 summarizes the main characteristics of the SpS.

Table 1. Main characteristics of the Spectrograph System (SpS) of the PFS instrument

| Overview | |
|---|---|
| Number of fibers | 2394 (~600 per module) |
| Location | 4$^{th}$ floor of the SUBARU Telescope (IR side) |
| **Optics** | |
| Wavelength range and Resolution | Blue: 0.38-0.65μm (R ≥ 2200) <br> Red: 0.63-0.97μm (R ≥ 2900) <br> Mid-res Red: 0.71-0.885μm (R ≥ 5000) <br> NIR: 0.94-1.26μm (R ≥ 4200) |
| Fibers core diameter | 128 μm |
| Fiber input F-ratio | F/2.8 (without focal ratio degradation) |
| Collimator focal ratio | F/2.5 |
| Camera focal ratio | F/1.1 |
| VPH Grating dimension | 280mm in diameter |
| Detector format | 4kx4k with 15μm pixel size |
| **Mechanics** | |
| Total Mass | <12 tons |
| Max. available volume | 5.3 x 6.3 x 2.7 m |
| **Operating conditions** | |
| Room temperature | +3°C |
| Altitude | 4200m above sea level |

## 2. CURRENT STATUS

The PFS collaboration, led by IPMU in Japan, consists of USP/LNA in Brazil, Caltech/JPL, Princeton, and JHU in USA, LAM in France, ASIAA in Taiwan, and NAOJ/Subaru.

Since the Conceptual Design Review (CoDR, phase A) in March 2012, the project has successfully passed two major milestones: the Preliminary Design Review (March 2013) and the Critical Design Review (CDR) held in March 2014.

Currently, the Spectrograph System (SpS) is in the implementation phase: the first spectrograph module will be integrated and tested at Laboratoire d'Astrophysique de Marseille (LAM) from mid-2015. The two first spectrograph modules will be delivered to SUBARU at the same time end-2016. The first light (with these two modules including all PFS elements) is foreseen in 2017.

# 3. ARCHITECTURE

The top-level Spectrograph System (SpS) architecture is given in Figure 1. The SpS is composed of four identical Spectrograph Modules (SM), which are constituted by five main units

- The Entrance Unit (ENU) aiming at capturing the input signal from fibers and feeding the cameras:
    - The beam enters each SM via the Fiber Slit Assembly (FSA), and is reflected by a collimating mirror (working at F/2.5). Two dichroics split the beam: the first dichroic reflects the "Blue" part of the spectrum and the second reflects the "Red" while the "NIR" is transmitted. Then the light is dispersed by three dedicated VPH gratings.
- The three Camera Units (one per waveband) aiming at detecting spectrum of the incoming light:
    - The spectrums are reimaged by three (almost) identical vacuum Schmidt cameras (one per channel) with a focal ratio of about F/1.1 on 4k x 4k detectors.
- The Local Control Unit (LCU) which regroups all the electronics and control boards.

The proposed architecture allows to keep the SpS modular in its design. This is required to allow for Assembly, Integration and Tests (AIT) and for its safe transport up to the summit. The four modules will be firstly integrated and fully validated at LAM. All sub-assemblies will be validated before their final integration on a module. Indeed, the sub-assemblies will be developed in parallel (in France but also in Brazil and in the USA), while the modules will be integrated in a sequence.

By its design, the SpS has a limited number of external interfaces. In particular, there is no direct optical interface with the telescope. Indeed the SpS will receive the light from the so called Cable B (fiber bundle coming from PFI).

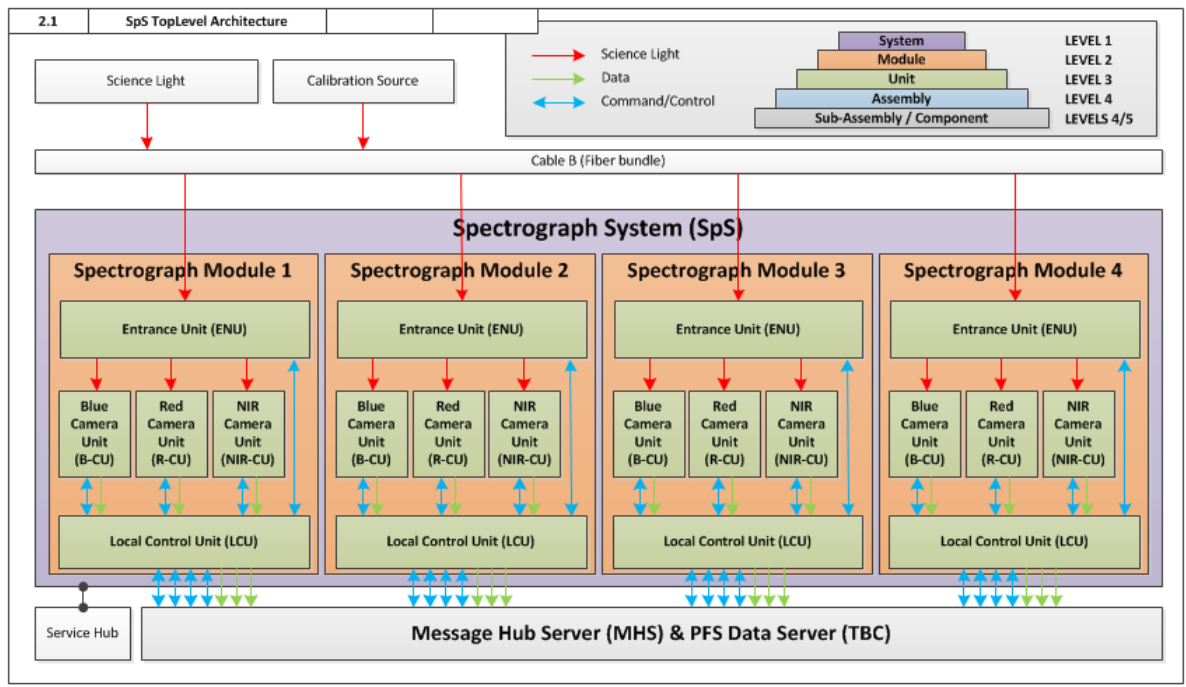

Figure 1. Spectrograph System (SpS) Architecture.

# 4. SPECTROGRAPH OPTICAL DESIGN

The optical architecture is based on a Schmidt collimator facing a Schmidt camera. It allows for high image quality with only few simple elements (high throughput) at the expense of the central obscuration, which leads to larger optics. The classical concept of the Schmidt camera has been modified by replacing the classical spherical mirror by a Mangin-like mirror (i.e meniscus lens with a reflective surface on the rear side of the glass).

The object plane is made of about 600 optical fibers arranged along a curved slit, coming out from the telescope primary focus. These fibers are collimated by a standard Schmidt chamber. Two dichroics are inserted in the beam (to form the three spectral channels, also referred as arms) between the collimator mirror and the Schmidt correctors (one per channel). The three Schmidt correctors are identical except for the coating, which is optimized for each waveband.

Once the linear object is collimated, the beam is dispersed by a VPHG assembly, one for each arm, in the perpendicular direction with respect to the object direction. All gratings are made on 320mm square by 20mm thick substrates and have a cover plate of the same size and thickness, both made of BK7. See [3] for more details on the VPHG (including the measurements made on full size prototypes).

For each of the three channels, a camera images the dispersed line on a square 4k x 4k detector (CCDs from Hamamatsu for the Blue and the Red, and a HgCdTe detector from Teledyne for the NIR). Each camera is based on a Mangin-Schmidt concept adapted to the fast focal ratio (F/1.09) and contains just four optical elements: a two-element refractive corrector, a Mangin mirror, and a field flattening lens. This simple design produces very good imaging performance (see Section 7) considering the wide field and wavelength range, and it does so in large part due to the use of a Mangin mirror (a lens with a reflecting rear surface) for the Schmidt primary.

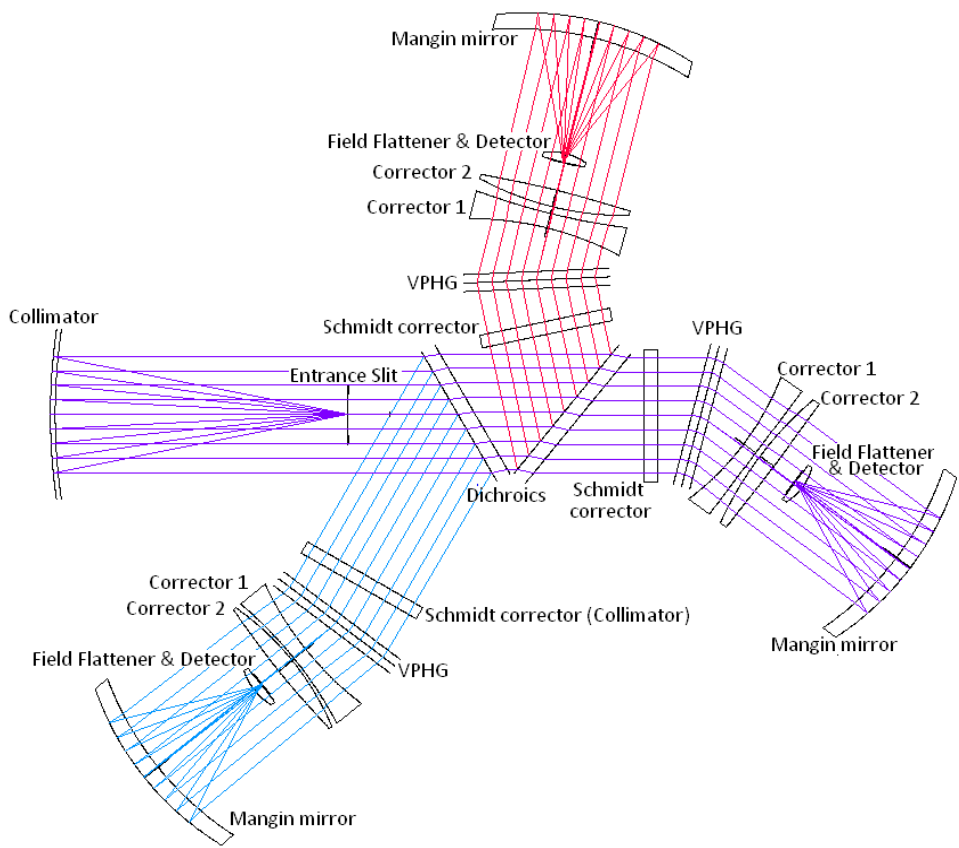

Figure 2. Optical layout of one spectrograph module. Each channel is color-coded: blue rays for the Blue channel, red rays for the Red channel, and purple rays for the NIR channel.

The proposed optical design has been optimized to achieve the requested image quality while simplifying the manufacturing of the whole optical system. The three channels are arranged to reduce the footprint of the spectrograph module while offering enough accessibility to the optical elements.

The Medium Resolution configuration is implemented on the red channel for a wavelength range of [0.710-0.885] microns with a resolving power of about 5000. The optical design consists in replacing the low resolution grating by a grism (i.e. a grating with two cemented prisms on both faces made of a high-index glass). The use of a grism allows changing the resolution (i.e. the dispersive element) while preserving the aiming of the optical axis i.e. without having to move the camera.

The optical layout is shown in Figure 2. More details on the optical design are given in [4].

## 5. MECHANICAL IMPLEMENTATION

The candidate location for setting four spectrograph modules has changed by decision of Subaru observatory from the third floor of the infrared instrument side (IR3) to the fourth floor of the same side (IR4), based on discussion in 2014 January Subaru Users' meeting and on recommendation from Subaru Advisory Committee.

Figure 3 shows the different components of one Spectrograph Module. They are mounted on a 1.9 x 2.4m optical bench made in CRFP. The optical bench for each spectrograph module is made of carbon fiber reinforcement plastic (CFRP), which has a low Coefficient of Thermal Expansion. This bench offers a high stability of spectrograph optics configuration against temperature variation. A modular cover will be installed over all elements in order to prevent stray light (both internal and external) and dust from impacting the optics, hence the performance.

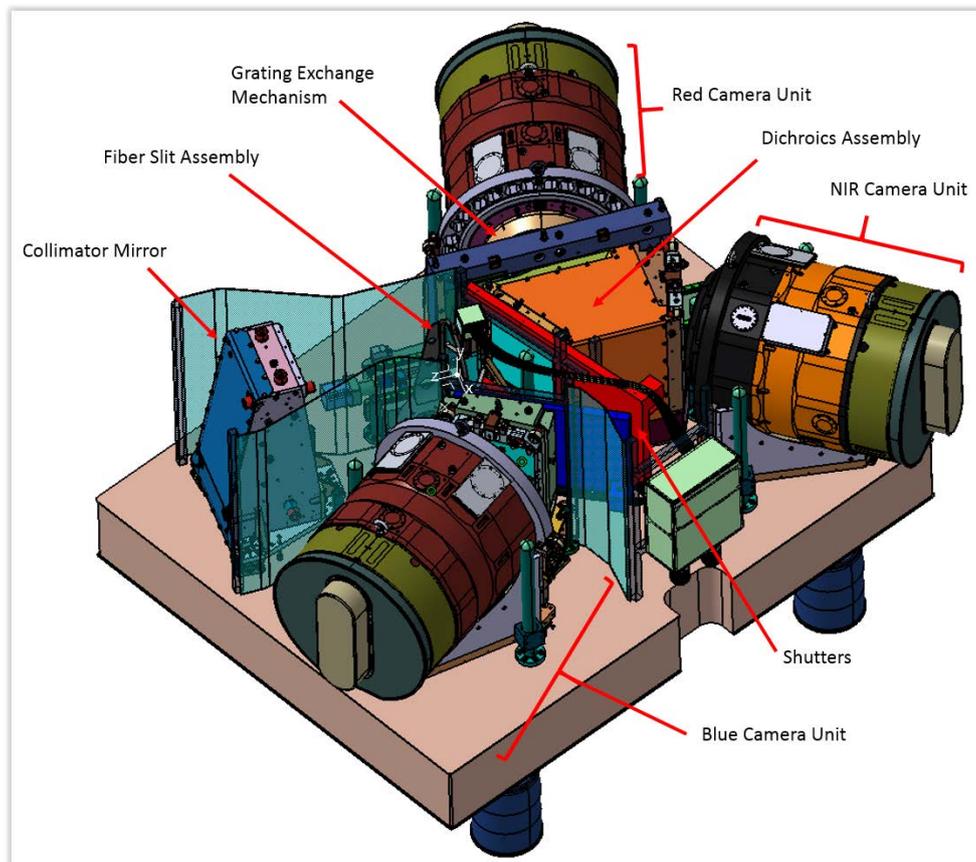

Figure 3. Spectrograph Module mechanical overview. The baffling covering the bench is not drawn.

The proposed mechanical design architecture deals with the need for modularity and repeatability. All elements on the optical bench are indexed to allow their accurate repositioning to cope with the several mounting and dismounting of the elements and in particular to minimize the activities at SUBARU. Furthermore, when firstly aligned at LAM, each element position will be verified and referenced with a dedicated metrology.

The Fiber Slit Assembly FSA is responsible to hold the slit head in place while limiting the obscuration of the optical beam. The slit head holds the 600 optical fibers along a curved slit. Two nickel masks (with holes) are used to position and aim the fibers.

The Fiber Slit Assembly will be mounted on a commercial hexapod. This mechanism offers the several advantages: it will be firstly used for the initial alignment of the slit (all degrees of freedom are required). Then in operation, the hexapod will be used to move the slit either along the optical axis (through focus to find the optical focus) or perpendicularly to the optical axis to perform flat fields.

Each spectrograph requires 2 shutters (different in size): a smaller one for the blue arm and a larger one shared by the red and NIR channels. The shutters (one-sided) apertures are 300mm and 400mm square. They run always horizontally with a closing/opening transient time < 1 sec guaranteed.

The VPHG for the red camera and the mid resolution grism are mounted on a linear exchange mechanism.

The two VIS cameras (Blue and Red) are identical from the mechanical point of view. The camera optics and detector are packaged in a cryostat and cooled by two Stirling cycle cryocoolers. The first corrector element serves as the vacuum window. The cameras are athermalized in the range [0; +20]°C allowing for alignment at ambient temperature.

The design of the cameras was mainly driven by the following considerations:

- Decouple the optics from vibrations generated by the cryocoolers. Indeed, they vibrate with a fundamental frequency of 60 Hz, with a non-negligible amount still in the second harmonic. Hence the mounts should be stiff with natural frequencies in excess of 120 Hz.
- Cool-down the detectors (and the optics in case of the NIR camera) while keeping all elements thermally isolated from the dewar walls. The VIS cameras only require one cryocooler, while the NIR requires two.
- Relax alignment tolerances for the optics (which are quite large and heavy) while minimizing the number of required adjustments.

The SpS is also involved in the metrology of the PFI positioners. Indeed, the fibers (located at the prime focus of the telescope) will be back illuminated from each Spectrograph Module. This is performed by illuminating a diffusing screen (located on one shutter blade) facing the collimating mirror and feeding the fibers.

## 6. MANUFACTURING AND AIT PHILOSOPHY

Optical components manufacture, assembly and performance verification are sub-contracted to an industry partner. The alignment of the Entrance Unit (ENU) and the pre-alignment of the Blue and Red camera will be also sub-contracted to the industry partner (except for the focal plane assembly – FPA). LAM will assemble and align the detector inside the Blue and Red camera. All alignments will be performed at ambient conditions while the performance tests of each module will be performed under cold environment at LAM.

It is a key driver to minimize AIT at SUBARU. Yet, there will be a minimum number of operations required to re-assemble the units (camera units, assemblies of the entrance units) onto the optical bench for each of the spectrograph module (SM), align them onto their mechanical references and verify the performance.

The optical alignments that will be required at SUBARU are the location of the slit with respect to the three camera focal planes and the orientation of the VPHG (roll) and the camera (tip/tilt) in order to align the grating lines with the slit (perpendicular) and the detector pixels matrix.

More details on the AIT are given in [5]

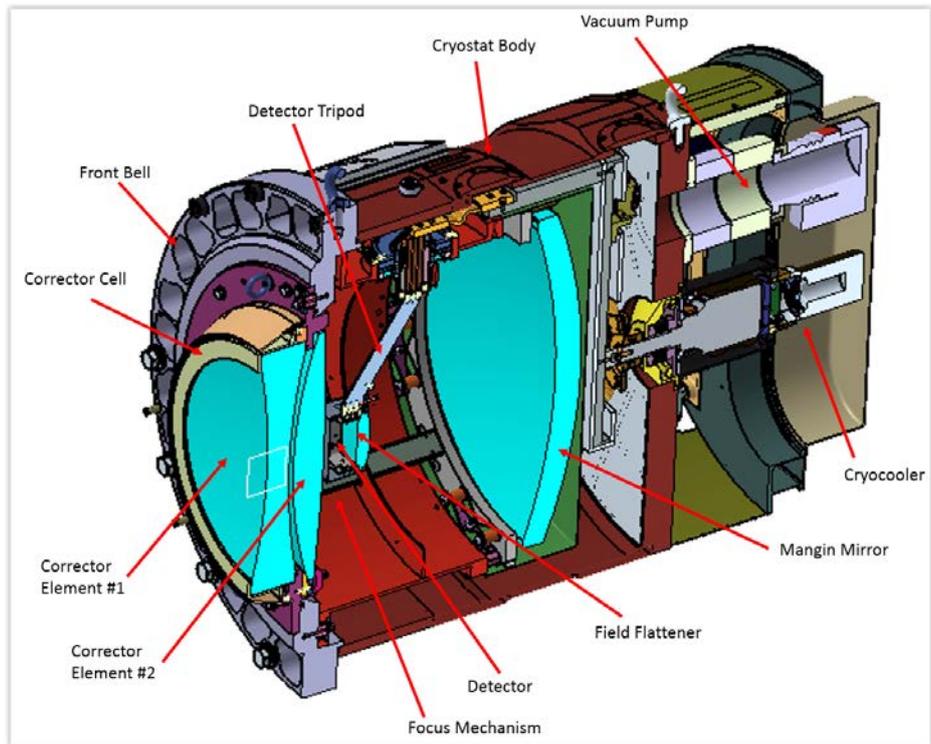

Figure 4. Opto-mechanical design of the VIS camera.

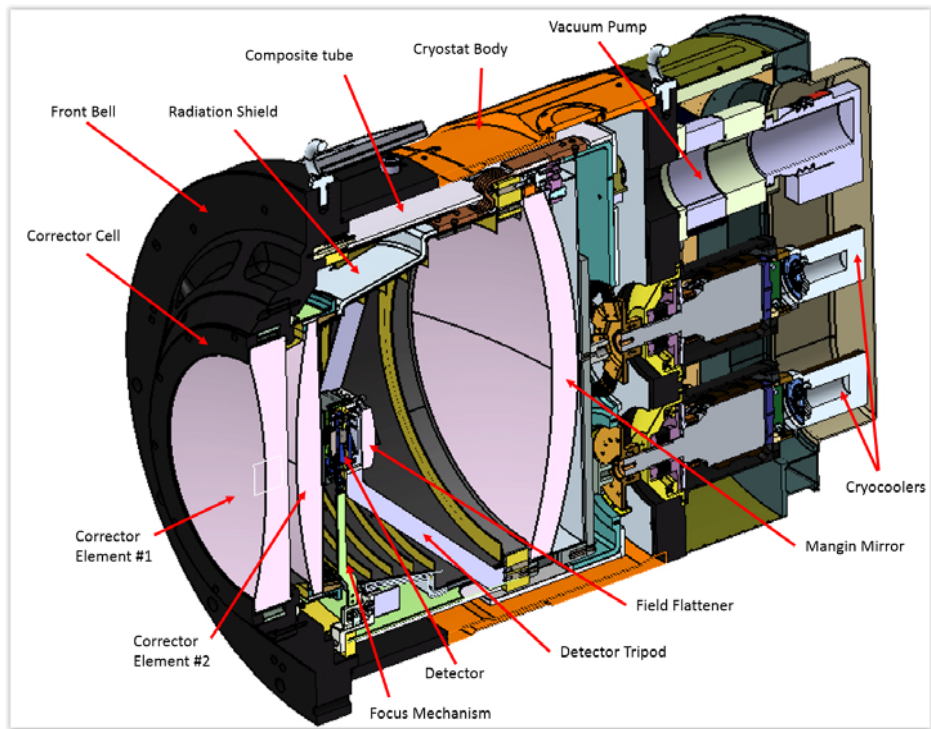

Figure 5. Opto-mechanical design of the NIR camera.

## 7. EXPECTED OPTICAL PERFORMANCE

The image quality is specified in terms of Extended Ensquared Energy EEE i.e. the Ensquared Energy considering the fiber as extended source. Then the EEE for one fiber shall be:

- $\geq 50\%$ within a square of 3 pixels for each spectral band in more than 95% of the detector area;
- $\geq 90\%$ within a square of 5 pixels for each spectral band in more than 95% of the detector area.

Figure 6 shows the distribution of the square size (in pixels) containing respectively 50% (left) and 90% (right) of Extended Ensquared Energy (for the different field points and wavelengths). This is obtained for the nominal theoretical design, and it appears that the theoretical design is well within the specifications. The tolerance design (including manufacturing and alignment errors) meets also the specifications and is described in [4].

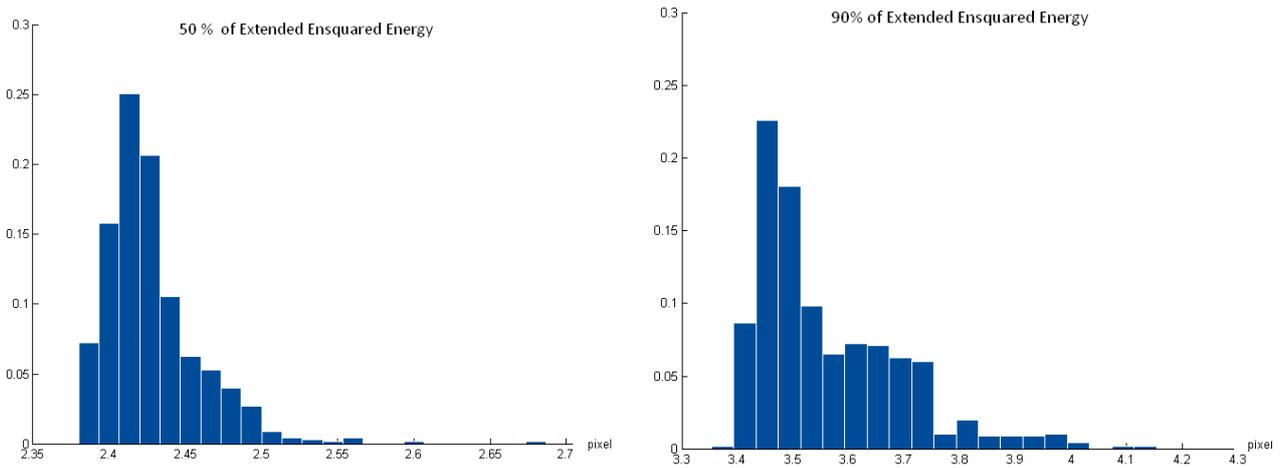

Figure 6. Distribution of the square size (in pixels) containing 50% (left) and 90% of EEE (right) in the three channels.

The next figure presents the matrix spot diagrams (matrix field versus wavelength) for each channel. The circles represent the image of one fiber on the detector (about 50microns).

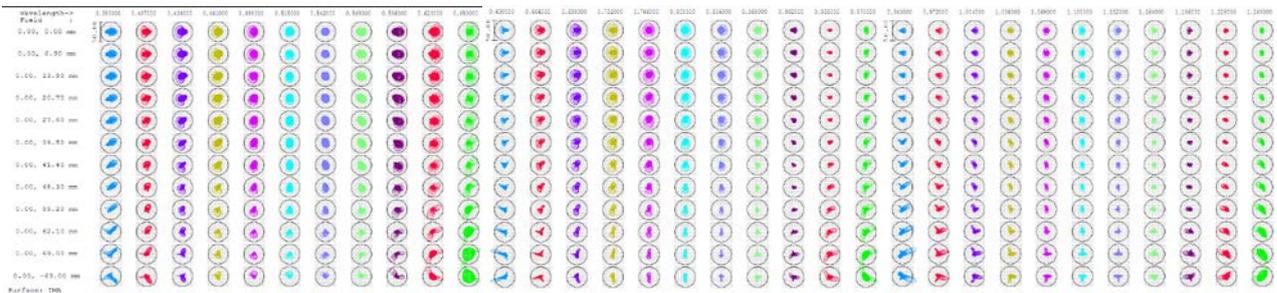

Figure 7. Matrix Spot Diagram for the three channels: Blue (left), Red (middle), and NIR (right).

More details on the optical design and the optical performance (in particular the relationship between RMS spot size and EEE) can be found in [4].

## 8. CONCLUSION

The Spectrograph System (SpS) has successfully passed Critical Design Review (CDR) in March 2014. It is currently in the construction phase. The first optical elements are under manufacturing: all Silica glass blanks are ready and being polished by Winlight System. Three prototype Volume Phase Holographic (VPH) gratings have been produced by Kaiser and have been tested [3].

The technical first light is planned in 2017.

## ACKNOWLEDGMENTS

We gratefully acknowledge support from the Funding Program for World-Leading Innovative R&D in Science and Technology (FIRST), program: "Subaru Measurements of Images and Redshifts (SuMIRe)", CSTP, Japan.